\documentclass[reqno,11pt]{article}
\usepackage{psfig, amsmath, amstexnb, amsthm}

\newtheorem{theorem}{Theorem}
\newtheorem{cor}{Corollary}

\theoremstyle{definition}
\newtheorem{assump}{Assumption}
\newtheorem{notation}{Notation}

\makeatletter

\def\enumerate{\ifnum \@enumdepth >3 \@toodeep\else
        \advance\@enumdepth \@ne 
        \edef\@enumctr{enum\romannumeral\the\@enumdepth}\list
        {\csname label\@enumctr\endcsname}
        {\setlength{\topsep}{1mm}
        \setlength{\parsep}{0mm}
        \setlength{\itemsep}{0mm}
        \setlength{\labelsep}{2mm}
        \settowidth{\leftmargin}{M.}
        \addtolength{\leftmargin}{\labelsep}
        \usecounter{\@enumctr}
        \def\makelabel##1{\hss\llap{##1}}}\fi}

\def\itemize{\ifnum \@itemdepth >3 \@toodeep\else \advance\@itemdepth \@ne
        \edef\@itemitem{labelitem\romannumeral\the\@itemdepth}%
        \list{\csname\@itemitem\endcsname}{
        \setlength{\topsep}{0mm}
        \setlength{\parsep}{0mm}
        \setlength{\parsep}{0mm}
        \setlength{\itemsep}{0mm}
        \setlength{\labelsep}{2mm}
        \settowidth{\leftmargin}{M.}
        \addtolength{\leftmargin}{\labelsep}
        \def\makelabel##1{\hss\llap{##1}}}\fi}

\def\captionheadfont@{\scshape}
\def\captionfont@{\small}
\long\def\@makecaption#1#2{%
  \setbox\@tempboxa\vbox{\color@setgroup
    \advance\hsize-3pc\noindent
    \captionfont@\captionheadfont@#1\@xp\@ifnotempty\@xp
        {\@cdr#2\@nil}{.\captionfont@\upshape\enspace#2}%
    \unskip\kern-3pc\par
    \global\setbox\@ne\lastbox\color@endgroup}%
  \ifhbox\@ne 
    \setbox\@ne\hbox{\unhbox\@ne\unskip\unskip\unpenalty\unkern}%
  \fi
  \ifdim\wd\@tempboxa=\z@ 
    \setbox\@ne\hbox to\columnwidth{\hss\kern-3pc\box\@ne\hss}%
  \else 
    \setbox\@ne\vbox{\unvbox\@tempboxa\parskip\z@skip
        \noindent\unhbox\@ne\advance\hsize-3pc\par}%
\fi
  \ifnum\@tempcnta<64 
    \addvspace\abovecaptionskip
    \moveright 1.5pc\box\@ne
  \else 
    \moveright 1.5pc\box\@ne
    \nobreak
    \vskip\belowcaptionskip
  \fi
\relax
}
\makeatother

\DeclareMathSymbol{\leqsymb}{\mathalpha}{AMSa}{"36}
\def\leqs{\;\leqsymb\;}
\DeclareMathSymbol{\geqsymb}{\mathalpha}{AMSa}{"3E}
\def\geqs{\;\geqsymb\;}
\DeclareMathSymbol{\gtreqqlesssymb}{\mathalpha}{AMSa}{"54}

\newcommand{\field}[1]{\mathbb{#1}}

\newcommand{\R}{\field{R}\,}	

\newcommand{\cA}{{\mathcal A}}	
\newcommand{\cC}{{\mathcal C}}	
\newcommand{\cD}{{\mathcal D}}	
\newcommand{\cN}{{\mathcal N}}	

\DeclareMathOperator{\e}{e}		
\DeclareMathOperator{\icx}{i}		
\DeclareMathOperator{\dd}{d}		
\DeclareMathOperator{\re}{Re}		

\def\dx#1{\dd\!#1}		
\def\defwd#1{{\sl#1}}		
\def\abs#1{\lvert#1\rvert}		
\def\norm#1{\lVert#1\rVert}		
\def\dtot#1#2{\frac{\dx{#1}}{\dx{#2}}}	
\def\sdtot#1#2{\dd\!_{#2}#1}		
\def\sdpar#1#2{\partial_{#2}#1}		
\def\fix#1{#1^{\star}}		

\def\eps{\varepsilon}
\def\sord{\approx}		
\def\nbh{neighbourhood}
\def\one{{\mathchoice {\rm 1\mskip-4mu l} {\rm 1\mskip-4mu l}
{\rm 1\mskip-4.5mu l} {\rm 1\mskip-5mu l}}}	

\def\brak#1{[#1]}			
\def\set#1{\{#1\}}			
\def\Order#1{{\mathcal O}(#1)}	

\def\bigbrak#1{\bigl[#1\bigr]}		
\def\bigpar#1{\bigl(#1\bigr)}		
\def\bigabs#1{\bigl|#1\bigr|}		
\def\bigsetsuch#1#2{\bigl\{#1\,\bigr|\bigl.\,#2\bigr\}}	

\def\bibtitle#1#2{#1, {\em #2}}                        
\def\bibfulltitle#1#2{#1, {\em #2}}                     
\def\bibref#1#2#3#4#5{#1 {\bf #2}  (#5), #3--#4}        
\def\bibarticle#1#2#3#4#5#6#7{\bibtitle{#1}{#2}, %
\bibref{#3}{#4}{#5}{#6}{#7}.}

\def\bibbook#1#2#3#4{#1, {\em #2} (#3, #4).}

\def\AMS{American Mathematical Society}
\def\AP{Ann.\ Physics}

\def\DE{Diff.\ Equ.}
\def\Dokl{Dokl.\ Akad.\ Nauk SSSR}

\def\JPA{J.\ Phys.\ A}

\def\NL{Nonlinearity}

\def\PRB{Phys.\ Rev.\ B}

\def\PRL{Phys.\ Rev.\ Letters}

\def\SAM{Stud.\ in Appl.\ Math.}
\def\SIAM{SIAM J.\ Appl.\ Math.}

\begin{document}



\title{Dynamic bifurcations: \\
hysteresis, scaling laws and feedback control}

\author{
N. Berglund\\
School of Physics, Georgia Institute of Technology\\
Atlanta GA 30332-0430, USA.}

\date{December 1, 1999}

\maketitle

\begin{abstract}
We review some properties of dynamical systems with slowly varying
parameters, when a parameter is moved through a bifurcation point of the
static system. Bifurcations with a single zero eigenvalue may create
hysteresis cycles, whose area scales in a nontrivial way with the adiabatic
parameter. Hopf bifurcations lead to the delayed appearance of
oscillations. Feedback control theory motivates the study of a bifurcation
with double zero eigenvalue, in which this delay is suppressed.
\end{abstract}


\section{Introduction}
\label{sec_1}

Many physical systems are described by ordinary differential equations
(ODEs) of the form 
\begin{equation}
\label{1:1}
\dtot{x}{t} = F(x,\lambda),
\end{equation}
where $x\in\R^n$ is a vector of \defwd{dynamic variables} and
$\lambda\in\R^p$ a set of \defwd{parameters}. When modeling the system by
the equation \eqref{1:1} we implicitly assume that the parameters are kept
constant, while only the dynamic variables change in time. There are,
however, situations in which the variables considered as parameters change
slowly in time, for instance:
\begin{itemize}
\item	parameters which may be difficult to control, such as the
temperature in a chemical reactor or the climate influencing an ecological
system;
\item	\defwd{control parameters}, such as the temperature difference in a
convection experiment or the supply voltage of an electrical device, which
are slowly varied in order to determine the bifurcation diagram
experimentally;
\item 	slowly time-dependent forcings such as a periodic magnetic field
acting on a magnet.
\end{itemize}
It is thus important to understand the relation between solutions of the
$p$-parameter family of autonomous ODEs \eqref{1:1}, and the time-dependent
system
\begin{equation}
\label{1:2}
\dtot{x}{t} = F(x,G(\eps t)),
\end{equation}
where the function $G(\tau):\R\to\R^p$ is given, $\eps$ is a small
parameter, and $\tau=\eps t$ is called \defwd{slow time}. It is convenient
to rewrite \eqref{1:2} as the singularly perturbed system
\begin{equation}
\label{1:3}
\eps\dtot{x}{\tau} = F(x,G(\tau)) = f(x,\tau).
\end{equation}
There is a large literature on equations of this type, see for instance the
textbooks \cite{MKKR,VBK}, the proceedings \cite{Benoit} and the paper
\cite{Jones} for reviews of different approaches. 

We will focus here on relations between the solutions of the slowly
time-dependent system \eqref{1:3} and the bifurcation diagram of the
``frozen'' system \eqref{1:1}. It turns out that the dynamics near
nonbifurcating equilibrium branches is relatively simple and well
understood. We summarize some of the relevant results in Section
\ref{sec_2}. Bifurcations, however, lead to new interesting phenomena, some
of which we will review in the next sections. 
\begin{itemize}
\item	Bifurcations in which a single eigenvalue becomes equal to zero may
lead to relaxation oscillations \cite{MR}, bifurcation delay and hysteresis
\cite{BK,B1}. In Section \ref{sec_3}, we present some results on the
$\eps$-dependence of the hysteresis area. 
\item 	The Hopf bifurcation, in which a pair of complex conjugate
eigenvalues crosses the imaginary axis, leads to the delayed appearance of
large amplitude oscillations. In Section \ref{sec_4} we present some of
Neishtadt's results  \cite{Ne,Ne2} on the determination of the bifurcation
delay.
\item	Finally, in Section \ref{sec_5}, we give a result on a bifurcation
problem with a double zero eigenvalue, which arises in the context of
feedback control \cite{B3,BS}.
\end{itemize}


\section{Non-bifurcating equilibria}
\label{sec_2}

If we naively replace $\eps$ by 0 in equation \eqref{1:3}, we obtain the
algebraic equation $f(x,\tau)=0$ which defines the equilibrium branches $x
= \fix{x}(\tau)$ of the frozen system. One can thus expect that for small
positive $\eps$, some solutions of the slowly time--dependent system stay
close to some equilibrium branch $\fix{x}(\tau)$. This can be made precise
in the following way.

We consider the equation
\begin{equation}
\label{2:1}
\eps\dtot{x}{\tau} = f(x,\tau)
\end{equation}
under the following assumption.

\begin{assump}
\label{ass2:1}
There exist an interval $I\in\R$, which need not be bounded, a
\nbh\ $\cD$ of the origin in $\R^n$, and positive constants $a_0,
w_0, d$ and $M$ such that the following properties are satisfied uniformly
in $\tau\in I$.  
\begin{itemize}
\item	The function $f(x,\tau):\cD\times I\to\R^n$ is of class $\cC^k$,
$k\geqs 2$.
\item 	There exists a function $\fix{x}(\tau):I\to\cD$ such that
$f(\fix{x}(\tau),\tau)=0$, with its derivative
$\sdtot{\fix{x}}{\tau}(\tau)$ bounded in norm by $w_0$.
\item 	The matrix $A(\tau) = \sdpar{f}{x}(\fix{x}(\tau),\tau)$ has
eigenvalues $a_j(\tau)$, $j=1,\dots,n$, such that $\abs{\re a_j(\tau)} \geqs
a_0 > 0$ for all $j$ (hyperbolicity).
\item	The function $b(y,\tau) = f(\fix{x}(\tau)+y,\tau) - A(\tau)y$ is
bounded by $M\norm{y}^2$ for $\norm{y}\leqs d$. 
\end{itemize}
\end{assump}

\begin{theorem}
\label{thm2:1}
Under Assumption \ref{ass2:1}, there exist strictly positive constants
$\eps_0$ and $c_1$, depending only on $a_0, w_0, d$ and $M$, such that if
$\eps\leqs\eps_0$:
\begin{enumerate}
\item	Equation \eqref{2:1} admits a particular solution $\bar{x}(\tau)$
with 
\begin{equation}
\label{2:2}
\norm{\bar{x}(\tau)-\fix{x}(\tau)}\leqs c_1\eps 
\qquad \forall\tau\in I.
\end{equation}
 
\item	If $k\geqs 3$, there exist functions $x_j(\tau)$, $j=1,\dots,k-2$
and a constant $c_{k-1}>0$ such that 
\begin{equation}
\label{2:3}
\norm{\bar{x}(\tau) - \brak{\fix{x}(\tau) + \sum_{j=1}^{k-2}
\eps^j x_j(\tau)}} \leqs c_{k-1}\eps^{k-1}
\qquad\forall\tau\in I.
\end{equation} 

\item	If $f$ is analytic in a complex \nbh\ of $\fix{x}(\tau)$,
there exist functions $x_j(\tau)$, $j\geqs 1$, constants $c, K>0$ and an
integer $N(\eps) = \Order{1/\eps}$ such that 
\begin{equation}
\label{2:4}
\norm{\bar{x}(\tau) - \brak{\fix{x}(\tau) + \sum_{j=1}^{N(\eps)}
\eps^j x_j(\tau)}} \leqs c \e^{-1/K\eps}
\qquad\forall\tau\in I.
\end{equation} 

\item	If $f$ is periodic in $\tau$ with period $T$, then $\bar{x}$ is also
periodic with period $T$.

\item	If all eigenvalues of $A(\tau)$ have a negative real part, there
exist constants $M_0,c_0,\kappa>0$ such that any solution of \eqref{2:1} with
initial condition such that $\norm{x(\tau_0)-\fix{x}(\tau_0)}\leqs c_0$ at
some $\tau_0\in I$ satisfies 
\begin{equation}
\label{2:5}
\norm{x(\tau)-\bar{x}(\tau)} \leqs M_0 \e^{-\kappa(\tau-\tau_0)/\eps} 
\norm{x(\tau_0)-\bar{x}(\tau_0)} 
\qquad \forall\tau\in I\cap[\tau_0,\infty).
\end{equation}

\item 	If $A$ has eigenvalues with both positive and negative real part,
there exist local invariant manifolds on which the motion is either
contracting or expanding.
\end{enumerate}
\end{theorem}

This result tells us that there exists indeed a particular solution tracking
the equilibrium $\fix{x}(\tau)$ at a distance of order $\eps$. This
particular solution is sometimes called \defwd{adiabatic} or \defwd{slow
solution}. It admits asymptotic series in $\eps$ which can be computed by
substitution into \eqref{2:1}. If the equilibrium is asymptotically stable,
the adiabatic solution attracts nearby solutions exponentially fast. In this
case, the relation
\begin{equation}
\label{2:6}
\lim_{\eps\to 0} x(\tau;\eps) = \fix{x}(\tau)
\qquad \forall\tau>\tau_0
\end{equation}
implies that we may indeed take the formal limit $\eps\to 0$ directly in
\eqref{2:1}.

Theorem \ref{thm2:1} has a long history. In the asymptotically stable case,
points 1.\ and 5.\ were originally proved in \cite{Tihonov,Grad}. A different
approach has been used in \cite{Fe}. The exponential bounds in the analytic
case are a result of an iterative scheme in \cite{Ne}, an alternative
approach can be found in \cite{Baesens}. The periodicity of solutions is a
consequence of the implicit function theorem, and the computation of
invariant manifolds in the hyperbolic case is explained in \cite{B1}. This
result can be extended to periodic orbits \cite{PR}.

Let us also point out that this result has an interesting consequence for
the linear equation 
\begin{equation}
\label{2:7}
\eps\dtot{y}{\tau} = A(\tau) y,
\end{equation}
which is similar to the Schr\"odinger equation as appearing in the
adiabatic theorem in quantum mechanics \cite{Berry,JKP}. This equation also
occurs if we linearize \eqref{2:1} around any particular solution.  The
eigenvalues of the matrix $A(\tau)\in\R^{n\times n}$ can be labeled by
continuous functions $a_1(\tau), \dots, a_n(\tau)$. 

\begin{assump}
\label{ass2:2}
There exists a partition into two groups $\set{a_1,\dots,a_p}$ and
$\set{a_{p+1},\dots,a_n}$ such that the \defwd{real gap}
\begin{equation}
\label{2:8}
\gamma = \inf_{\substack{\tau\in I \\ 1\leqs i\leqs p\\ p+1\leqs j\leqs
n}} \bigabs{\re\bigpar{a_i(\tau) - a_j(\tau)}},
\end{equation}
is strictly positive.
\end{assump}

\begin{theorem}[\cite{B1}]
\label{thm2:2}
Assume that $A(\tau)$ is of class $\cC^3$ for $\tau\in I$, and that
Assumption \ref{ass2:2} holds. For sufficiently small $\eps$
and $\tau\in I$, there exists an invertible matrix $S(\tau,\eps)$ such that
the change of variables $y = S(\tau,\eps)z$ transforms \eqref{2:7} into
\begin{equation}
\label{2:9}
\eps\dot{z} = D(\tau,\eps) z,
\end{equation}
where $D(\tau,\eps)$ is block--diagonal, with one block of size $p\times p$
and eigenvalues $a_j(\tau)+\Order{\eps}$ for $j=1,\dots,p$, and another
block of size $(n-p)\times (n-p)$
and eigenvalues $a_j(\tau)+\Order{\eps}$ for $j=p+1,\dots,n$. 
The matrices $S(\tau,\eps)$ and $D(\tau,\eps)$ can be expanded into powers 
of $\eps$, up to exponential order in the analytic case.  
\end{theorem}

Since the connection between this result and Theorem \ref{thm2:1} does not
seem to be well known, we sketch the proof in Appendix \ref{app_A}.

\begin{cor}
\label{cor2:1}
Assume that the eigenvalues of $A(\tau,\eps)$ have uniformly disjoint real
parts, that is, 
\begin{equation}
\label{2:11}
\inf_{\substack{\tau\in I \\ 1\leqs i< j\leqs n}} 
\bigabs{\re\bigpar{a_i(\tau) - a_j(\tau)}} > 0.
\end{equation}
Then equation \eqref{2:7} can be diagonalized by a change of variables $y
= S(\tau,\eps)z$, and thus its principal solution can be written in the form
\begin{equation}
\label{2:12}
U(\tau,\tau_0) = S(\tau,\eps)
\begin{pmatrix}
\e^{\alpha_1(\tau,\tau_0)/\eps} & & 0 \\
& \ddots & \\
0 & & \e^{\alpha_n(\tau,\tau_0)/\eps}
\end{pmatrix}
S(\tau_0,\eps)^{-1},
\end{equation}
where 
\begin{equation}
\label{2:13}
\alpha_j(\tau,\tau_0) = \int_{\tau_0}^{\tau} a_j(s) \dx s + \Order{\eps}.
\end{equation}
\end{cor}


\section{Single zero eigenvalue: hysteresis and scaling laws}
\label{sec_3}

Theorem \ref{thm2:1} breaks down if some eigenvalues of the linearization
$A(\tau) = \sdpar{f}{x}(\fix{x}(\tau),\tau)$ cross the imaginary axis, that
is, in the case of a bifurcation. In this case, new phenomena may occur. As
noted in \cite{MKKR}, the center manifold theorem can be applied to the
equation $\dtot{}{t}(x,\tau,\eps) = (f(x,\tau),\eps,0)$ to reduce the
dimension of the system to the number of bifurcating eigenvalues. 

The simplest case occurs when a single eigenvalue vanishes at the
bifurcation point. We write the reduced equation in the form
\begin{equation}
\label{3:1}
\eps\dtot{x}{\tau} = f(x,\tau), 
\qquad x\in\R.
\end{equation}
We shall assume that a stable equilibrium branch $\fix{x}(\tau)$, existing
for negative $\tau$, bifurcates at the origin. This requires that $f(0,0) =
\sdpar{f}{x}(0,0) = 0$. 

Let us consider a few examples:
\begin{enumerate}
\item	In the \defwd{saddle--node bifurcation}, the stable branch meets an
unstable branch, and no equilibrium exists for positive $\tau$. This
results in the adiabatic solution jumping to some other attractor. Jump
phenomena were first studied in \cite{Po}, and later in \cite{Hab,MR,MKKR}.
\item	In the \defwd{transcritical bifurcation}, a stable and an unstable
branch exchange stability. It has been shown in \cite{LS1} that
generically, solutions will track the outgoing stable branch, although it
occasionally happens that the unstable branch is tracked for some time
\cite{NS1}. 
\item	In the \defwd{pitchfork bifurcation}, the stable branch becomes
unstable, and two new stable branches are created. It has been shown in
\cite{LS2} that generically, solutions will track one of the two outgoing
stable branches, where the choice of the branch depends on the geometry. If
the original equilibrium does not depend on $\tau$, however, the
bifurcation may be delayed \cite{BK,NS1,EM}.
\end{enumerate}

\begin{figure}
 \centerline{\psfig{figure=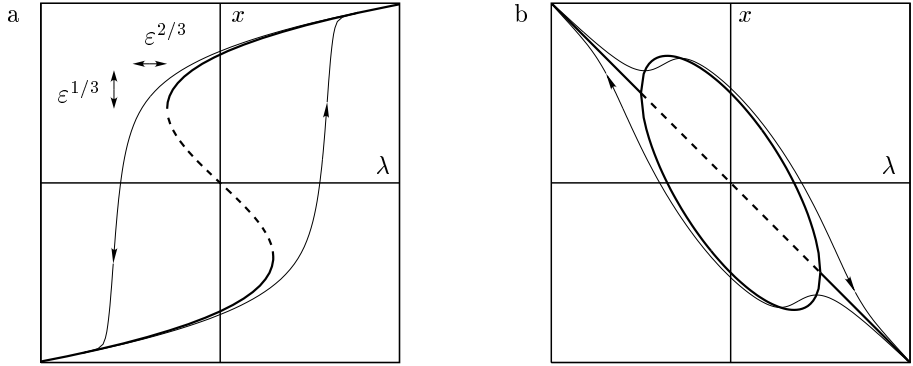,height=50mm,clip=t}}
 \caption[]
 {Comparison between solutions of equations with slowly moving parameters
 (light curves) and the bifurcation diagram of the associated frozen system
 (heavy curves, where full curves represent stable equilibria, and broken
 curves unstable ones). (a) The bifurcation diagram of equation \eqref{3:2}
 contains two saddle-node bifurcations. Solutions converge to a hysteresis
 cycle with area satisfying $\cA(\eps)-\cA(0)\sord\eps^{2/3}$. (b) The
 bifurcation diagram of equation \eqref{3:3} contains two pitchfork
 bifurcations. Solutions track the upper branch for increasing $\lambda$
 and the lower one for decreasing $\lambda$, creating a hysteresis cycle
 with area satisfying $\cA(\eps)-\cA(0)\sord\eps^{3/4}$.}
\label{fig:1}
\end{figure}

These local properties of bifurcations have interesting global consequences
if the parameter is varied periodically. The bifurcation points may be
arranged in such a way that different branches are tracked for increasing
and decreasing parameter, and thus there is \defwd{hysteresis} (and relation
\eqref{2:6} no longer holds). A classic
example is the equation
\begin{equation}
\label{3:2}
\eps\dtot{x}{\tau} = x - x^3 + \lambda,
\qquad \lambda = \sin\tau,
\end{equation}
which leads to a familiar hysteresis curve (Fig.~\ref{fig:1}a). The curves
$x(\tau)$ display so-called \defwd{relaxation oscillations} in which
slow and fast motions alternate. Another example is 
\begin{equation}
\label{3:3}
\eps\dtot{x}{\tau} = (x+\lambda) (1-\lambda^2-(x+\lambda)^2), 
\qquad \lambda = 2\sin\tau,
\end{equation}
see Fig.~\ref{fig:1}b. In combination with other global mechanisms, such as
inertia in the equation
\begin{equation}
\label{3:4}
\eps^2 \dtot{^2x}{\tau^2} = -\gamma\eps \dtot{x}{\tau} + \lambda(\tau)x -
x^3, \qquad \lambda = \sin\tau,
\end{equation}
hysteresis may also lead to a chaotic motion \cite{BK,BK2}.

Another interesting effect of bifurcations is that equilibrium branches are
no longer tracked at a distance of order $\eps$, but at a distance scaling
with some fractional power of $\eps$. This results in hysteresis areas
scaling in nontrivial ways with $\eps$ \cite{Jung,Hohl,GBS}. Similar
scaling laws have been observed in ferromagnets \cite[and references
therein]{RKP,SE}, and they continue to be a domain of active research.

We present here a method to determine scaling laws in an easy, geometrical
way \cite{BK,B1}. The key idea is to understand the behaviour of the variable
$y=x-\fix{x}(\tau)$, which satisfies an equation of the form
\begin{equation}
\label{3:5}
\eps\dtot{y}{\tau} = a(\tau)y + b(y,\tau) + \eps w(\tau),
\end{equation}
where $a(\tau)=\sdpar{f}{x}(\fix{x}(\tau),\tau)$,
$w(\tau)=-\sdtot{\fix{x}}{\tau}(\tau)$ and $b(y,\tau)=\Order{y^2}$. 

\begin{notation}
\label{not3:1}
We write $x(\tau,\eps)\sord y(\tau,\eps)$ if there exist constants
$c_{\pm}>0$, uniform in $\tau$ and $\eps$, such that $c_-y(\tau,\eps)
\leqs x(\tau,\eps) \leqs c_+ y(\tau,\eps)$ for small $\tau$ and $\eps$.
\end{notation}

\begin{assump}\hfill
\label{ass3:1}
\begin{itemize}
\item	The function $f(x,\tau)$ is of class $\cC^k$, $k\geqs 3$, for
$0\leqs x\leqs d$ and $\tau_0\leqs\tau<0$, and admits the Taylor expansion
\begin{equation}
\label{3:6}
f(x,\tau) = \sum_{n+m<k} c_{nm}x^n\tau^m + \sum_{n+m=k} R_{nm}(x,\tau)
x^n\tau^m, \qquad c_{00}=c_{10}=0.
\end{equation}
\item	For $\tau_0\leqs\tau<0$, there exists a function
$\fix{x}(\tau)\sord\abs{\tau}^q$ ($q>0$) such that
$f(\fix{x}(\tau),\tau)=0$ and $f(x,\tau)<0$ for $\fix{x}(\tau)<x\leqs d$.
\item	The function $a(\tau)=\sdpar{f}{x}(\fix{x}(\tau),\tau)$ satisfies
$a(\tau)\sord-\abs{\tau}^p$, where
\begin{equation}
\label{3:7}
p = \min_{n\geqs 1,m\geqs 0} 
\bigsetsuch{q(n-1)+m}{\text{($n+m<k$ and $c_{nm}\neq0$) or $n+m=k$}}.
\end{equation}
\end{itemize}
\end{assump}

The numbers $q$ and $p$ can be determined geometrically by \defwd{Newton's
polygon}, which is defined as the convex envelope of all points in
$(n,m)$-plane such that either $n+m<k$ and $c_{nm}\neq0$, or $n+m=k$. Then
$-q$ is the slope of a segment of the polygon, and $p$ is the ordinate at 1
of this segment. The condition \eqref{3:7} is generically satisfied, as can
be seen by inserting $\fix{x}(\tau) \sord\abs{\tau}^q$ into the Taylor
expansion \eqref{3:6}. 

\begin{theorem}[\cite{B1}]
\label{thm3:1}
Under Assumption \ref{ass3:1}, any solution starting sufficiently close to
$\fix{x}(\tau_0)$ reaches the $\Order{\eps}$-\nbh\ of
$\fix{x}(\tau)$ at a time $\tau_1 = \tau_0 + \Order{\eps\abs{\ln\eps}}$ and
satisfies
\begin{equation}
\label{3:8}
x(\tau) - \fix{x}(\tau) \sord
\begin{cases}
\dfrac{\eps}{\abs{\tau}^{p+1-q}} 
&\text{for $\tau_1 \leqs \tau \leqs -\eps^{1/(p+1)}$} \\
\eps^{q/(p+1)} 
&\text{for $-\eps^{1/(p+1)} \leqs \tau \leqs 0$.} 
\end{cases}
\end{equation}
The area between $x(\tau)$ and $\fix{x}(\tau)$ scales as 
\begin{equation}
\label{3:9}
\cA(\eps) \sord 
\begin{cases}
\eps^{(q+1)/(p+1)} 
&\text{if $q<p$,} \\
\eps\abs{\ln\eps} 
&\text{if $q=p$,} \\
\eps 
&\text{if $q>p$.}
\end{cases}
\end{equation}
\end{theorem}

We sketch the proof in Appendix \ref{app_B}. 
It is more difficult to study the behaviour of solutions after the
bifurcation, where there is no such general result. One can, however, obtain
much information by studying the dynamics of the distance between $x$ and
the various equilibrium branches emerging from the bifurcation point, using
similar methods.

For the branch of the pitchfork bifurcation with vertical tangent, we have
$q=\frac12$ and $p=1$, which implies that this branch is tracked at a
distance of order $\eps^{1/4}$, and the area scales as $\eps^{3/4}$.
For the saddle-node bifurcation, we have $q=p=\frac12$. The branch is
tracked at a distance of order $\eps^{1/3}$, and the jump turns out to be
delayed by a time scaling as $\eps^{1/(p+1)}=\eps^{2/3}$, which accounts for
the scaling law in Fig.~\ref{fig:1}a.


\section{Hopf bifurcation and bifurcation delay}
\label{sec_4}

We consider now the equation \eqref{2:1} in the case where a pair of
complex conjugate eigenvalues crosses the imaginary axis from left to
right. The frozen system generically displays one of the following two
behaviours:
\begin{itemize}
\item	in the \defwd{supercritical Hopf bifurcation}, a stable periodic
orbit is created,
\item	in the \defwd{subcritical Hopf bifurcation}, an unstable periodic
orbit is destroyed.
\end{itemize}

\begin{figure}
 \centerline{\psfig{figure=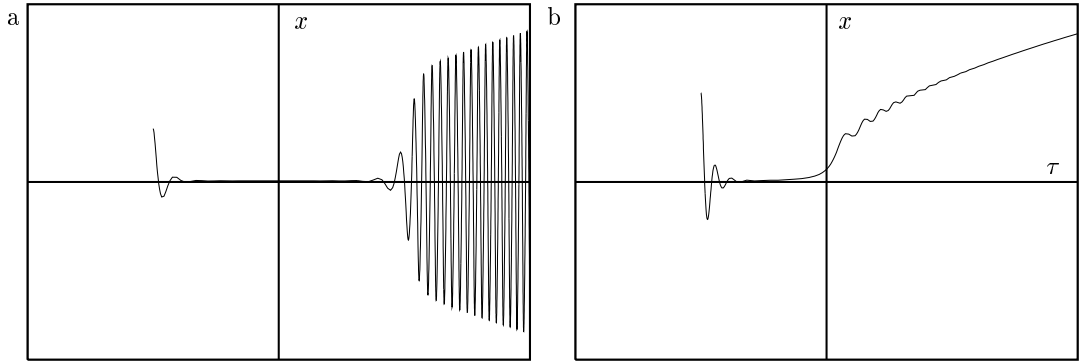,height=50mm,clip=t}}
 \caption[]
 {(a) The slow passage through a Hopf bifurcation (occurring at the origin)
 leads to the delayed appearance of oscillations. The delay depends on the
 initial condition, but admits a saturation value, called buffer time. (b)
 The delay is suppressed by an appropriate feedback control, which creates
 a bifurcation with double zero eigenvalue. Solutions track a stable
 equilibrium branch emerging from the bifurcation point.}
\label{fig:2}
\end{figure}

When the parameter is slowly varied through the bifurcation point, one
would expect the appearance of oscillations with a slowly increasing
amplitude in the first case, and a jump transition in the second case.
This, however, does not occur if the system is smooth enough: the moment at
which the trajectory departs from the unstable equilibrium turns out to be
delayed with respect to the moment of the bifurcation (Fig.~\ref{fig:2}a).
This \defwd{bifurcation delay} was first described in \cite{Sh}, analysed
rigorously by Neishtadt \cite{Ne,Ne2}, and observed in various physical
systems \cite{BER,HE}.  We now state a version of Neishtadt's result on the
computation of the delay. 

\begin{assump}
\label{ass4:1}
There exist an open interval $I\in\R$ containing 0 and a \nbh\
$\cD$ of the origin in $\R^n$ such that
\begin{itemize}
\item	The function $f(x,\tau)$ is analytic in a complex \nbh\ of
$I\times\cD$.
\item	There is a curve $\fix{x}(\tau):I\to\R^n$ with
$f(\fix{x}(\tau),\tau)=0$.
\item	The matrix $A(\tau) = \sdpar{f}{x}(\fix{x}(\tau),\tau)$ has two
eigenvalues $a(\tau)\pm\icx\omega(\tau)$, where $a(0)=0$, $a'(0)>0$,
$\omega(0)\neq 0$, and $a(\tau)$ has the same sign as $\tau$; all other
eigenvalues of $A(\tau)$ have a strictly negative real part.
\end{itemize}
\end{assump}

For $\tau$ in a complex \nbh\ of $I$, we can define the function
\begin{equation}
\label{4:1}
\Psi(\tau) = \int_0^\tau \brak{a(s)+\icx\omega(s)}\dx s. 
\end{equation}
For negative $\tau_0\in I$, we define the function 
\begin{equation}
\label{4:2}
\Pi(\tau_0) = \sup_{\tau>\tau_0} \bigsetsuch{\tau}{\re\Psi(s) <
\re\psi(\tau_0), \tau_0<s<\tau}.
\end{equation}
We have $\Pi(\tau_0)>0$ for $\tau_0<0$ because $a(s)$ is negative for
negative $\tau$, which implies that $\re\Psi(\tau)$ is decreasing for
negative $\tau$. Moreover, we have $\lim_{\tau_0\to 0-}\Pi(\tau_0) = 0$ and
$\lim_{\tau_0\to 0-}\Pi'(\tau_0) = -1$. 
For instance, if $a(\tau)=\tau$, then $\Pi(\tau_0)=-\tau_0$. 

\begin{theorem}[\cite{Ne}]
\label{thm4:1}
For any strictly negative $\tau_0\in I$ there exist constants $\eps_0, c_0,
c_1 > 0$ and a continuous function $r(\eps)$ with
$\lim_{\eps\to0}r(\eps)=0$ such that for $\eps<\eps_0$, any solution with
initial condition such that $\norm{x(\tau_0) - \fix{x}(\tau_0)}\leqs c_0$
satisfies
\begin{equation}
\label{4:3}
\norm{x(\tau) - \fix{x}(\tau)} \leqs c_1\eps 
\qquad\text{for $\tau_0 + r(\eps) \leqs \tau \leqs
\min\set{\tau_+,\Pi(\tau_0)} - r(\eps)$.}
\end{equation}
The so-called buffer time $\tau_+>0$ depends only on the behaviour of the
eigenvalue $a(\tau)+\icx\omega(\tau)$ in the complex plane.
\end{theorem}

The computation of the buffer time $\tau_+$ is explained in \cite{Ne2}, see
also \cite{Benoit}. Roughly speaking, it is the largest real time which can
be connected to the negative real axis by a path with constant $\re\Psi$.
It is usually of the order of $\omega(0)$. The remarkable fact is that
unlike for the pitchfork bifurcation, the bifurcation delay exists also when
the equilibrium $\fix{x}(\tau)$ depends on $\tau$. 

\section{Double zero eigenvalue and feedback control}
\label{sec_5}

Jumps and large amplitude oscillations may have catastrophic consequences
for a device undergoing bifurcations. This has lead some researchers to
design feedback controlled systems. A simple affine control system takes
the form
\begin{equation}
\label{5:1}
\dtot{x}{t} = f(x,\lambda) + b\,u(x,\lambda),
\end{equation}
where $f(x,\lambda)$ is the uncontrolled vector field satisfying Assumption
\ref{ass4:1}, $b\in\R^n$ is a given vector, which defines the direction in
which the system can be steered, and $u(x,\lambda)\in\R$ is the scalar
feedback control. In previous works \cite{Ae,Ab,MS}, feedback controls have
been designed in order to make the bifurcation supercritical, so that
orbits departing from the unstable equilibrium have a bounded amplitude.

The results of the previous section show however that in the case of a
dynamic Hopf bifurcation, a jump transition occurs even in the
supercritical case. To avoid such a jump, one may try to suppress the
bifurcation delay by moving the imaginary parts of the bifurcating
eigenvalues to zero, in order to push the buffer time to zero. This will
produce a bifurcation with double zero eigenvalue. Moreover, the control
should ensure that the system is nonlinearly stable at the bifurcation
point, in such a way that stable equilibria are created which attract the
solutions.

In \cite{B3} we construct a feedback control which satisfies these
requirements. We then show that the dynamics can be reduced to the
two-dimensional equation
\begin{align}
\label{5:2}
\eps\dtot{x}{\tau} &= y \\
\nonumber
\eps\dtot{y}{\tau} &= \mu(\tau)x + 2a(\tau)y + \gamma(\tau) x^2 +
\delta(\tau) xy  - x^2y - x^3  
 + \Order{(x^2+y^2)^2} + \eps R(x,y,\tau,\eps),
\end{align}
where $a(\tau)$ is the same as in Assumption \ref{ass4:1}, the functions
$\mu(\tau), \gamma(\tau)$ and $\delta(\tau)$ vanish at $\tau=0$ and
$R(0,0,\tau,0)$ is directly related to the drift of $\fix{x}(\tau)$. The
linearization of \eqref{5:2} at the origin has eigenvalues
$a(\tau)\pm\mu(\tau)^{1/2}$. The frozen version of this system is a
codimension-four unfolding of the singular vector field $(y,- x^2y - x^3)$
which has been studied in detail \cite{KKR}.

\begin{theorem}[\cite{B3}]
\label{thm5:1}
Assume that the right-hand-side of \eqref{5:2} is analytic, $a'(0)>0$ and
$\mu'(0) > 0$.  There exist positive constants $d$, $T$, $M$, $\kappa$ and
a \nbh\ $\cN$ of the origin in $\R^2$ with the following property. For
every  $\tau_0\in [-T,0)$, there is a constant $c_1>0$ such that for
sufficiently small $\eps$, any solution of \eqref{5:2} with initial
condition $(x,y)(\tau_0)\in\cN$  satisfies
\begin{align}
\label{5:3}
\abs{x(\tau)} &\leqs M \eps\abs{\tau}^{-1}, & 
\abs{y(\tau)} &\leqs M \eps\abs{\tau}^{-1/2}, &
&\text{$\tau_1(\eps) \leqs \tau \leqs 
-(\eps/d)^{2/3}$,} \\ 
\abs{x(\tau)} &\leqs M \eps^{1/3}, & 
\abs{y(\tau)} &\leqs M \eps^{2/3}, &
&\text{$-(\eps/d)^{2/3} \leqs \tau \leqs 
(\eps/d)^{2/3}$},  
\end{align}
where $\tau_1(\eps)=\tau_0+c_1\eps\abs{\ln\eps}$.
If, moreover, the relations  $\mu'(0) > 2 a'(0)$ and $R(0,0,0,0) \neq 0$
hold, then for $(\eps/d)^{2/3} \leqs \tau \leqs T$ we have
\begin{equation}
\label{5:5}
\begin{split}
\abs{x(\tau)-x_+(\tau)} &\leqs M \bigbrak{\eps\tau^{-1} +
\eps^{1/2}\tau^{-1/4}\e^{-\kappa\tau^2/\eps}}, \\ 
\abs{y(\tau)} &\leqs M \bigbrak{\eps\tau^{-1/2} +
\eps^{1/2}\tau^{1/4}\e^{-\kappa\tau^2/\eps}}, 
\end{split}
\end{equation}
where 
\begin{equation}
\label{5:6}
x_+(\tau) = 
\begin{cases}
\sqrt{\mu} + \Order{\tau}, & 
\text{if $R(0,0,0,0) > 0$,} \\
-\sqrt{\mu} + \Order{\tau}, & 
\text{if $R(0,0,0,0) < 0$}
\end{cases}
\end{equation} 
are equilibria of \eqref{5:2}, i.e., the right-hand side of \eqref{5:2}
vanishes when $x=x_+, y=0$ and $\eps=0$. 
\end{theorem}

This behaviour is illustrated in Fig.~\ref{fig:2}b. The important fact is
that instead of tracking the unstable equilibrium (represented by the
origin) for some time, the trajectory smoothly tracks the equilibrium
$x_+(\tau)$, a feature that can be used to detect the fact that a
bifurcation has occurred. The behaviour is more complicated if
$\mu'(0)<2a'(0)$ \cite{B3,BS}. 


\section*{Acknowledgments}

The research presented in Sections \ref{sec_2} and \ref{sec_3} was done in
collaboration with Herv\'e Kunz, and partly supported by the Swiss National
Fund. Results in Section \ref{sec_5} were obtained in collaboration with
Klaus Schneider, and supported by the Nonlinear Control Network of the
European Community, Grant ERB FMRXCT--970137.


\appendix
\section{Proof of Theorem \ref{thm2:2}}
\label{app_A}

If $A(\tau)$ is of class $\cC^k$ and satisfies Assumption \ref{ass2:2},
there exists a matrix $S_0(\tau)$ of class $\cC^k$ such that
$S_0^{-1}AS_0=D_0$ is block-diagonal \cite{Wasow,Bel}. The change of
variables $y=S_0(\tau)y_0$ yields the equation
\begin{equation}
\label{A1:1}
\eps\dtot{y_0}{\tau} = A_0(\tau)y_0,
\qquad
A_0(\tau;\eps) = S_0^{-1}AS_0 - \eps S_0^{-1}\dtot{S_0}{\tau} = 
\begin{pmatrix}
A_{11} & \eps A_{12} \\
\eps A_{21} & A_{22}
\end{pmatrix}.
\end{equation}
This equation can be transformed into the block-diagonal form \eqref{2:9} by
the change of variables $y_0 = S_1(\tau)z$ if we {\em impose} that $S_1$
satisfy the differential equation
\begin{equation}
\label{A1:2}
\eps\dtot{S_1}{\tau} = A_0(\tau)S_1 - S_1 D(\tau).
\end{equation}
We will look for a solution with matrices of the form
\begin{equation}
\label{A1:3}
S_1(\tau) = 
\begin{pmatrix}
\one_p & \eps S_{12} \\
\eps S_{21} & \one_{n-p}
\end{pmatrix},
\qquad
D(\tau) = 
\begin{pmatrix}
D_1 & 0 \\ 0 & D_2
\end{pmatrix},
\end{equation}
where $S_{12}\in\R^{p\times(n-p)}$ and $S_{21}\in\R^{(n-p)\times p}$. 
Substitution into \eqref{A1:2} leads to
\begin{equation}
\label{A1:4}
\eps\dtot{S_{12}}{\tau} = A_{12} + A_{11}S_{12} - S_{12}A_{22} -
\eps^2S_{12}A_{21}S_{12}, 
\qquad
D_2 = A_{22} + \eps^2 A_{21}S_{12}
\end{equation}
and similar relations for $S_{21}, D_1$. Let $L$ be the linear operator $L:
X\mapsto A_{11}X - X A_{22}$. For $\eps=0$, the eigenvalues of $L$ are
exactly given by $a_i - a_j$, $1\leqs i\leqs p$, $p+1\leqs j\leqs n$
\cite{Wasow, Bel}.
Assumption \ref{ass2:2} implies that $L$ is invertible (its inverse can be
represented as a complex integral \cite{Krein}), and by the implicit function
theorem, the right-hand-side of \eqref{A1:4} vanishes for $S_{12} =
\fix{S}(\tau) = -L^{-1}A_{12} + \Order{\eps}$. The Fr\'echet derivative of
\eqref{A1:4} around $\fix{S}$ is equal to $L+\Order{\eps^2}$, and thus, by
Assumption \ref{ass2:2}, $\fix{S}$ is hyperbolic for small $\eps$. Theorem
\ref{thm2:1} can then be applied to show the existence of a solution
$S_{12}(\tau) = \fix{S}(\tau) + \Order{\eps}$. \qed


\section{Proof of Theorem \ref{thm3:1}}
\label{app_B}

The assertion on $\tau_1$ is a consequence of Theorem \ref{thm2:1}.   Let
$\mu=\frac{q}{p+1}$ and $\nu=\frac{1}{p+1}$.   We first consider equation
\eqref{3:5} for $\tau_1\leqs\tau\leqs -\eps^\nu$. By assumption, we have
$a(\tau)\leqs -a_+\abs{\tau}^p$ for some $a_+>0$. Let $b_0>0$. Since
$y(\tau_1)=\Order{\eps}$, there exists by continuity a time
$\fix{\tau}\in(\tau_1,0]$ such that $y(\tau)\leqs a_+\abs{\tau}^q/2b_0$ for
$\tau_1\leqs\tau\leqs\fix{\tau}$. Moreover, we may assume that either
$\fix{\tau}=-\eps^\nu$ or $y(\fix{\tau}) = a_+\abs{\tau}^q/2b_0$. Using the
Tayor expansion of $f$, one can show that for $\tau\in[\tau_1,\fix{\tau}]$,
$b(y,\tau)\leqs b_0 y^2 \abs{\tau}^{p-q}$, provided $b_0$ is large enough.
Then for $\tau_1\leqs\tau\leqs\fix{\tau}$ 
\begin{equation}
\label{A2:1}
\eps\dtot{y}{\tau} \leqs -\frac{a_+}{2b_0} \abs{\tau}^p y + \eps w_0
\abs{\tau}^{q-1}.
\end{equation}
This linear equation can be solved explicitly, to prove that $y(\tau)\leqs
c_+\eps\abs{\tau}^{q-p-1}$ on the time interval under consideration, for
some $c_+>0$. Let us write $\fix{\tau}=-\eps^{\nu-\delta}, \delta\geqs 0$.
Then $y(\fix{\tau})\leqs c_+\eps^{\delta(p+1)}\abs{\fix{\tau}}^q$. By
definition of $\fix{\tau}$, we have either $\fix{\tau}=-\eps^\nu$, or
$y(\fix{\tau}) = a_+\abs{\fix{\tau}}^q/2b_0$, and then
$\eps^{\delta(p+1)}\geqs a_+/2b_0c_+$, which shows that $\fix{\tau}\geqs
-(2b_0c_+/a_+)^\nu \eps^\nu \sord -\eps^\nu$. One obtains a lower bound in
the same way, which proves that $y(\tau)\sord\eps\abs{\tau}^{q-p-1}$ for
$\tau_1\leqs\tau\leqs\fix{\tau}\sord-\eps^\nu$.

For $\fix{\tau}\leqs\tau\leqs 0$, the rescaling $y=\eps^\mu z$, $\tau =
\eps^\nu \sigma$ gives
\begin{equation}
\label{A2:2}
\dtot{z}{\sigma} = \tilde{a}(\sigma)z + \tilde{b}(z,\sigma) +
\tilde{w}(\sigma),
\end{equation}
where $\tilde{a}(\sigma) = \eps^{\nu-1}a(\eps^{\nu}\sigma) \sord
-\abs{\sigma}^p$, $\tilde{w}(\sigma) = \eps^{\nu-\mu}w(\eps^{\nu}\sigma)
\sord \abs{\sigma}^{q-1}$, and one can show with the Taylor expansion  
that $\tilde{b}(\sigma,z) = \eps^{\nu-\mu-1}b(\eps^{\mu}z,
\eps^{\nu}\sigma)$ is smaller than a constant which does not depend on
$\eps$. Using the fact that $\sdtot{z}{\sigma} \geqs -a_-\abs{\sigma}^p z -
b_0 z^2$, we obtain by solving this Bernoulli equation that $z$ is bounded
below by a positive constant uniform in $\eps$. Moreover, the hypothesis
$f(\fix{x}(\tau)+y,\tau)<0$ for $y>0$ shows that $\tilde{a}(\sigma)z +
\tilde{b}(z,\sigma) < 0$ for $z>0$, which yields the upper bound
$\sdtot{z}{\sigma}\leqs w_0\abs{\sigma}^{q-1}$. This shows that $z(\sigma)$
is also bounded from above by a constant independent of $\eps$. This proves
that $y(\tau)\sord\eps^\mu$ for $\fix{\tau}\leqs\tau\leqs 0$.	\qed


\end{document}